\documentclass[aps,pre,preprint,superscriptaddress,nofootinbib]{revtex4-1}

\usepackage{graphicx}
\usepackage{amsmath}
\usepackage{amssymb}
\usepackage{bm}

\begin{document}

\title{A structural approach to relaxation in glassy liquids}
\author{S. S. Schoenholz$^\ddagger$}
\email{schsam@sas.upenn.edu}
\affiliation{Department of Physics, University of Pennsylvania, Philadelphia, Pennsylvania 19104, USA}
\author{E. D. Cubuk$^\ddagger$}
\affiliation{Department of Physics and School of Engineering and Applied Sciences, Harvard University, Cambridge, Massachusetts 02138, USA}
\author{D. M. Sussman}
\affiliation{Department of Physics, University of Pennsylvania, Philadelphia, Pennsylvania 19104, USA}
\author{E. Kaxiras}
\affiliation{Department of Physics and School of Engineering and Applied Sciences, Harvard University, Cambridge, Massachusetts 02138, USA}
\author{A. J. Liu}
\email{ajliu@sas.upenn.edu}
\affiliation{Department of Physics, University of Pennsylvania, Philadelphia, Pennsylvania 19104, USA}
\date{\today}
\let\thefootnote\relax\footnotetext{$\ddagger$\hspace{1pc}Equal contribution.}
\begin{abstract}
When a liquid freezes, a change in the local atomic structure marks the transition to the crystal. When a liquid is cooled to form a glass, however, no noticeable structural change marks the glass transition.  Indeed, characteristic features of glassy dynamics that appear below an onset temperature, $T_0$,~\cite{angell95,sastry98,debenedetti01} are qualitatively captured by mean field theory~\cite{parisi10,berthier11,charbonneau14}, which assumes uniform local structure at all temperatures.  Even studies of more realistic systems have found only weak correlations between structure and dynamics~\cite{gilman75,berthier07,royall08,manning11,jack14}.  This raises the question: is structure important to glassy dynamics in three dimensions?  Here, we answer this question affirmatively by using machine learning methods to identify a new field, ``softness?" which characterizes local structure and is strongly correlated with rearrangement dynamics. We find that the onset of glassy dynamics at $T_0$ is marked by the onset of correlations between softness (i.e. structure) and dynamics.  Moreover, we use softness to construct a simple model of slow glassy relaxation that is in excellent agreement with our simulation results, showing that a theory of the evolution of softness in time would constitute a theory of glassy dynamics.
\end{abstract}

\maketitle

To look for correlations between structure and dynamics, one typically tries to find a quantity that encapsulates the important physics, such as free volume, bond orientational order, locally preferred structure, etc.  In contrast to this approach, we use a machine learning method designed to find a structural quantity that is strongly correlated with dynamics. Earlier, we applied this approach to the simpler problem of classifying particles as being ``soft'' if they are likely to rearrange or ``hard'' otherwise~\cite{cubukschoenholz15}. We describe a particle's local structural environment with $M=166$ ``structure functions"~\cite{behler07} that respect the overall isotropic symmetry of the system and include radial density and bond angle information. We then define an $M$-dimensional space, $\mathbb R^M$, with an orthogonal axis for each structure function. The local structural environment of a particle $i$ is thus encoded as a point in $M$-dimensional space. We assemble a ``training set" from molecular dynamics simulations consisting of equal numbers of ``soft" particles that are about to rearrange and ``hard" particles have not rearranged in a time $\tau_\alpha$ preceding their structural characterization, and find the best hyperplane separating the two groups using the support vector machines (SVM) method~\cite{libsvm,SVM}. Finally, we define the \emph{softness} $S_i$, of particle $i$ as the shortest distance between its position in $\mathbb R^M$ and the hyperplane, where $S_i>0$ if $i$ lies on the soft side of the hyperplane and $S_i<0$ otherwise. 

We study a 10,000-particle 80:20 bidisperse Kob-Andersen Lennard-Jones glass~\cite{kob94} in $d=3$ at different densities $\rho$ and temperatures $T$ above its dynamical glass transition temperature. All results here are for particles of species $A$ only. However, the results are qualitatively the same for particles of both species. At each density we select a training set of $6,000$ particles, taken from a molecular dynamics trajectory at the lowest $T$ studied, to construct a hyperplane in $\mathbb R^M$. We then use this hyperplane to calculate $S_i(t)$ for each particle $i$ at each time $t$ during an interval of $30,000\tau$ at each $\rho$ and $T$. 

We can deduce the most important structural features contributing to softness either by training on fewer structure functions or by examining the projection of the hyperplane normal onto each orthogonal structure function axis.  Both analyses yield a consistent picture (see supplementary information): the most important features are the density of neighbors at the first peaks of the radial distribution functions $g_{AA}(r)$ and $g_{AB}(r)$; these two features alone give 77\% prediction accuracy for rearrangements.  Particles with more neighbors at the first peaks of $g(r)$ have a lower softness, and are thus more stable. These results are reminiscent of the cage picture, in which an increase of population in the first-neighbor shell suppresses rearrangements, or the free-volume picture, in which particles whose surroundings are closely-packed are more stable than those with more loosely-packed neighborhoods~\cite{berthier11_2}. Overall, soft particles typically have a structure that is more similar to a higher-temperature liquid, where there are more rearrangements, while hard particles whose structure appears closer to a lower-temperature liquid~\cite{percus58}.

Fig.~1 (a) is a snapshot with particles colored according to their softness. Evidently, $S$ has strong spatial correlations. Fig.~1 (b) shows the distribution of softness, $P(S)$, and the distribution of softness for particles just before they go through a rearrangement, $P(S|R)$. We see that 90\% of the particles that undergo rearrangements have $S>0$. We have also tested other sets of structure functions (see supplemental information) and found nearly identical accuracy. Softness is therefore a highly accurate predictor of rearrangements that is reasonably robust to the set of structure functions chosen.

\begin{figure}
\label{fig1}
\centering
\includegraphics[width=0.8\textwidth]{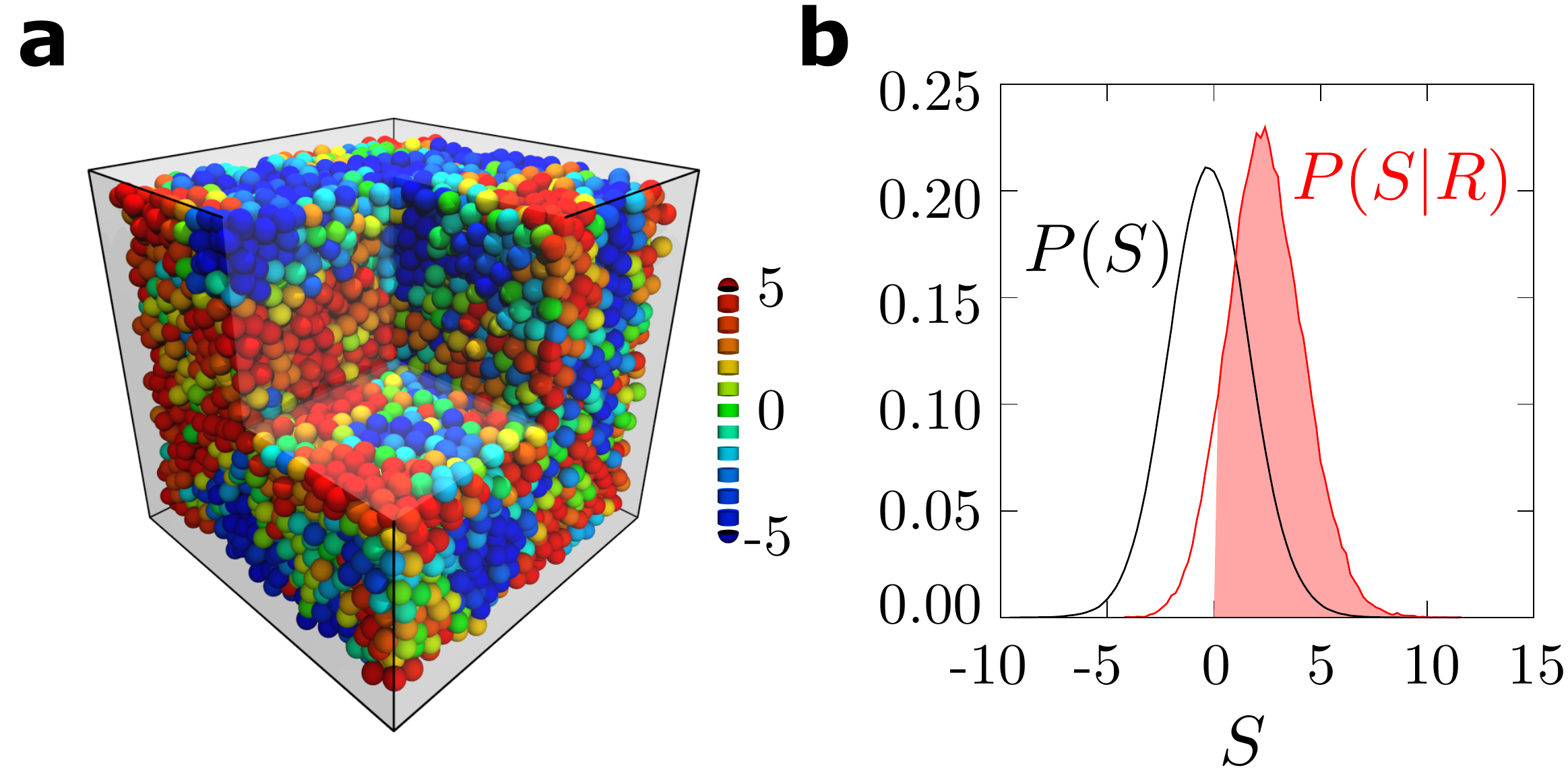}
\caption{\textbf{The characteristics of the softness field. a}, A snapshot of the system at $T=0.47$ and $\rho = 1.20$ with particles colored according to their softness from red (soft) to blue (hard). \textbf{b}, The distribution of softness of all particles in the system (black) and of those particles that are about rearrange (red). 90\% of the particles that are about to rearrange have $S > 0$ (shaded region). None of the data included in this plot were in the training set.}
\end{figure}

We next show that the probability that particles rearrange is a function of their softness. This probability is calculated as the fraction of particles of a softness, $S$, that are rearranging at a given time, $P_R(S)$. We plot $P_R(S)$ in Fig.~2 (a) in solid lines at temperatures ranging from $T=0.47$ (blue) to $T=0.58$ (red). At each $T$ we see that $P_R(S)$ is a strong function of softness, increasing by several orders of magnitude, especially  at the lower temperatures, in the range $S= -3$ to $S=3$. A similar, albeit more modest, relationship was seen in~\cite{smessaert14}. When $P_R(S)$ is plotted as a function of $1/T$ for several values of softness, Fig.~2 (b), the probability that a particle of softness $S$ will rearrange has Arrhenius behavior,  $P_R(S)=P_0(S)\exp(-\Delta E(S)/T)$ where $P_0(S)$ and $\Delta E(S)$ depend on $S$. Confirming this observation, $P_R(S)/P_0(S)$ collapses over many orders of magnitude for all temperatures when plotted against $\Delta E(S)/T$, as shown in the inset of Fig.~2(b).

\begin{figure}[ht!]
\centering
\includegraphics[width=0.8\textwidth]{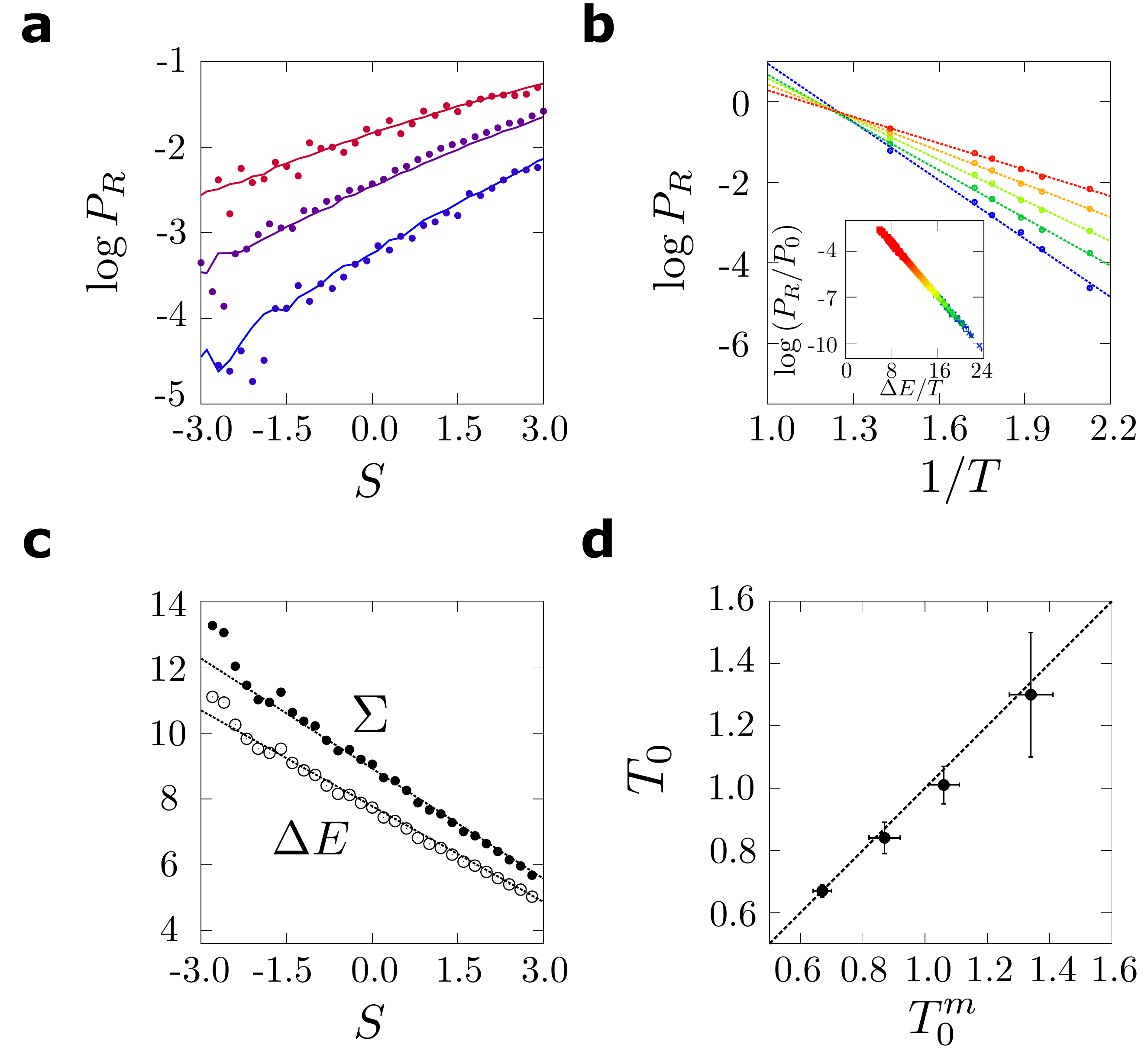}
\caption{\textbf{The relationship between softness and dynamics. a}, 
The probability that particles rearrange as a function of their softness, $P_R(S)$, for temperatures $T=$0.47, 0.53, and 0.58 plotted in blue to red. Solid lines are measurements from molecular dynamics trajectories (solid lines). Dashed lines present the probability computed using the Arrhenius form for $P_R(S)$ (dashed lines). Points represent the probabilities calculated from the zero-time derivative of the overlap, $-dq(S,t)/dt$ at $T=0.47$ and $T=0.58$. \textbf{b}, $P_R(S)$ as a function of $1/T$ for 5 different softness values from $S\sim -3$ (blue) to $S\sim 3$ (red). The inset shows the collapse of these probabilities when $P_R/P_0$ is plotted against $\Delta E/T$. \textbf{c}, $\Delta E$ and $\Sigma$, where $P_R(S)=\exp(\Sigma-\Delta E/T)$, vs.~softness $S$. \textbf{d}, predicted onset temperature $T_0$ vs.~$T^m_0$, onset temperature measured by Keys, et al.~\cite{keys11}, for densities $\rho=1.15,1.20, 1.25, 1.30$. The straight line corresponds to $T_0=T_0^m$. }   
\label{fig2}
\end{figure}

An Arrhenius form emerges when a kinetic process depends on a single energy scale $\Delta E(S)$.  In Fig.~2 (c) we plot $\Delta E(S)$ and $\Sigma(S) \equiv \ln P_0(S)$ vs.~$S$. Both terms depend nearly linearly on $S$: $\Delta E=e_0-e_1 S$ and $\Sigma=\Sigma_0-\Sigma_1 S$ where all four coefficients are positive and independent of $T$. Our results are consistent with the interpretation that at low temperatures, harder regions of the glassy liquid with higher energy barriers are frozen out while softer regions are not, leading to heterogeneous dynamics.  These heterogeneities smooth out with increasing temperature, and vanish altogether once $P_R(S)$ no longer depends on softness.  This occurs at the temperature $T_0$ where the softness dependence of $\Sigma$ exactly cancels that of $\Delta E/T_0$ and so $T_0=e_1/\Sigma_1$. This result can also be seen visually in Fig.~2 (b) where the different Arrhenius predictions for $P_R(S)$ all intersect at a single temperature, $T_0$, where the probability of rearrangement will be independent of softness. In Fig.~2 (d) we compare our prediction for $T_0$ to the onset temperature of glassy dynamics measured by Keys \textit{et al.}~\cite{keys11}, $T_0^m$, at different densities. The excellent agreement between the predicted $T_0$ and the measured values implies that the onset of glassy dynamics at $T=T_0$ coincides with the onset of correlations between structure (softness) and dynamics.

\begin{figure}
\centering
\includegraphics[width=0.8\textwidth]{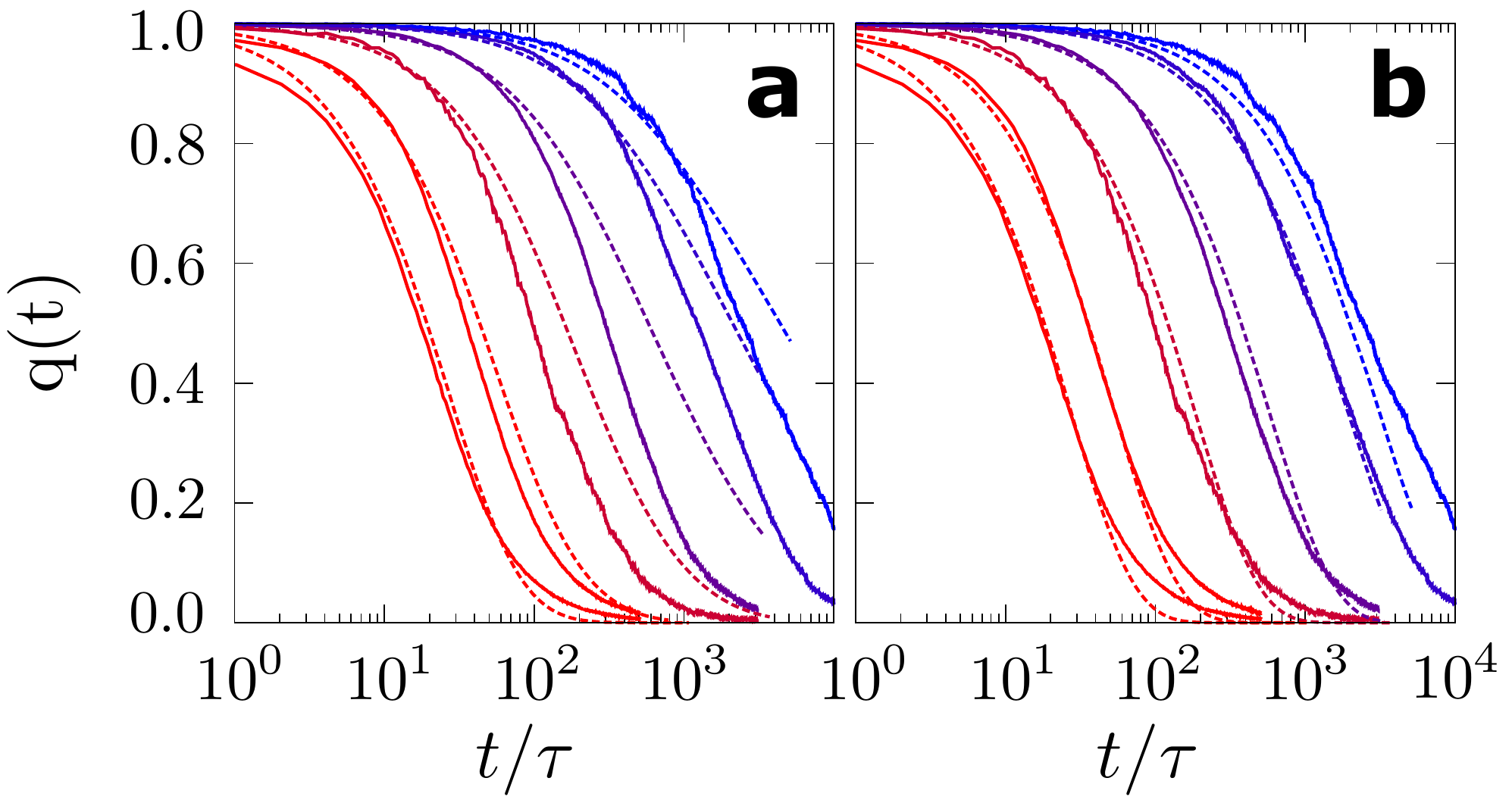}
\caption{\textbf{Overlap calculated from softness a}, Solids lines are the measured overlap function, for temperatures $T=$ 0.45, 0.47, 0.53, 0.58, 0.63, and 0.70, from blue to red, respectively. The dashed lines in the insets show predictions assuming each Arrhenius process is independent of one another. \textbf{b}, The solid lines in the insets are the same as in \textbf{a}.  Dashed lines are predictions for the overlap function from $P_R(S)$ including changes in the softness field induced by spatial correlation between rearranging particles.}
\label{fig3}
\end{figure}

We explore next the relationship between softness and the nonexponential decay of the overlap function 
$$q(t)= \frac 1N\sum_i\Theta(|r_i(t)-r_i(0)| - a)$$ 
where $N$ is the number of particles in the system, $r_i$ is the position of particle $i$, and $\Theta$ is the Heaviside function. We take $a=0.5$~\cite{keys07}. In Fig.~3 (a)-(b) we plot the overlap function for different temperatures at $\rho=1.20$. Our aim is to understand the form of the decay of $q(t)$ from the behavior of the rearrangement probability, $P_R(S)$.  To begin, we define the
contribution to the overlap from particles whose softness was initially $S$ at $t=0$, $q(S,t)$. The total overlap is $q(t) = \int dS  q(S,t)P(S)$. Because $q(S,t)$ is the fraction of particles with initial softness $S$ that have not rearranged after a time $t$, we expect $\frac{dq(S,t)}{dt}|_{t=0} = -c_a P_R(S)$ (see Supplementary Information for details) where $c_a$ is the fraction of rearrangements that displace particles by more than $a$.
This is indeed the case, as is evident from the data in Fig.~2 (a), when $\frac{dq(S,t)}{dt}|_{t=0}$ (points) is overlaid with $P_R(S)$ (solid lines). 

If we now assume that each particle rearranges with probability $P_R(S)$ as an independent Arrhenius process according to Fig.~2, then we can predict the decay of $q(S,t)$ using a simple discrete model: it can be written in terms of the probability that a particle of softness $S$ does not rearrange for $t-1$ timesteps before finally rearranging at time $t$, $(1-P_R(S))^{t-1}P_R(S)$. The resulting prediction for $q(t)$ (dashed) is shown in Fig.~3 (a) for several different temperatures. While the prediction is not poor, its accuracy decreases at longer times, particularly at lower temperatures. 

We now show that the discrepancy between our naive theory and the decay of $q(t)$ primarily results from a crucial neglected feature: 
even if a given particle does not rearrange, 
its local structural environment --and therefore its softness-- can be altered if nearby particles rearrange. This physics is reminiscent of facilitation.

\begin{figure}
\centering
\includegraphics[width=0.9\textwidth]{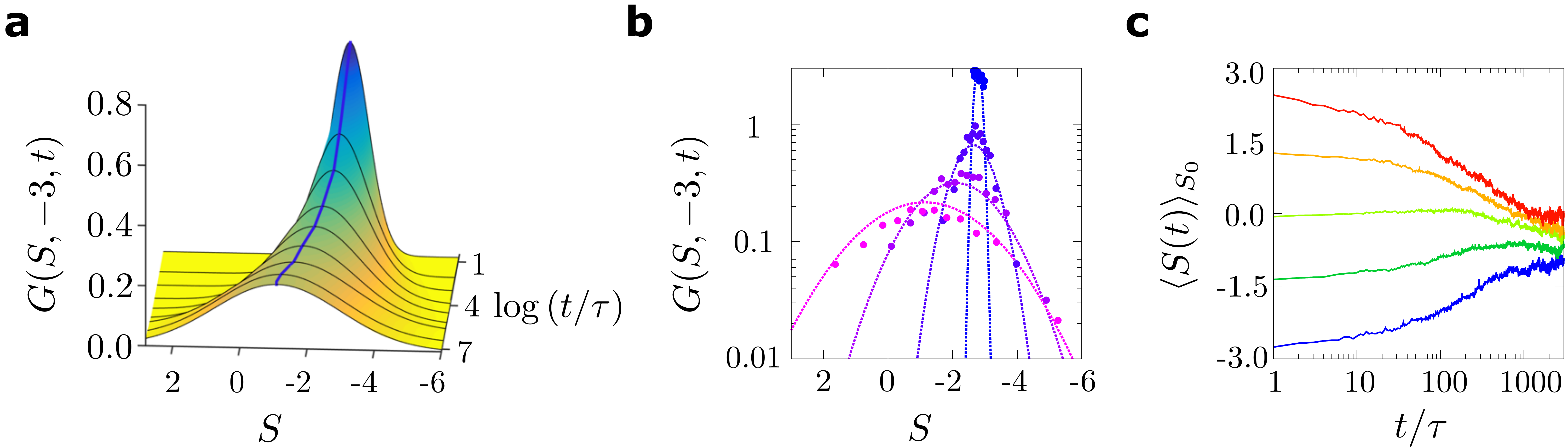}
\caption{\textbf{Time evolution of softness a}, The stochastic evolution of softness in time as seen in through the evolution of the Gaussian approximation to the distribution of softness. \textbf{b}, The time evolution of the softness distribution for a collection of particles with initial softness $S_0\sim-3$ from $t=0$ (blue) to $t=1000\tau$ (pink). Points are the measured histogram values, and the dashed lines are Gaussian approximations to the distribution. \textbf{c}, The time evolution of the average softness for particles that start from several softness values ranging from $S_0 \sim -3$ (blue) to $S_0 \sim 3$ (red).
}
\label{fig4}
\end{figure}

To take this facilitation into account, we calculate the ``softness propagator",  $G(S,S_0,t)$, 
that is, the distribution of softness at time $t$ for particles that start with a softness $S_0$ at $t=0$ and move less than a distance $a$ after a time $t$ 
({\textit{i.e.} that do not rearrange in a time $t$). 
Fig.~4 (a) shows a Gaussian approximation to $G(S,S_0=-3,t)$. We see that $G(S,S_0,t)$ is sharply peaked around $S_0$ at small $t$ but widens and shifts with increasing $t$ reminiscent of directed diffusion. Fig.~4 (b) shows $G(S,S_0=-3,t)$ at several different times, where points are measured probabilities and dashed lines are their Gaussian approximations. In Fig.~4 (c) we plot the mean softness evaluated as $\langle S(t)\rangle_{S_0} = \int dS SG(S,S_0,t)$ for several different values of $S_0$. For each $S_0$ the average softness of particles evolves towards the mean of the equilibrium softness distribution over a time period of approximately $\tau_\alpha$.  The softness propagator is evaluated only for particles that have not rearranged, so Fig.~4 shows that rearrangements of nearby particles affect a particle's softness significantly.

Our first naive prediction based on the assumption that particles rearrange independently corresponds to $G(S,S_0,t) = \delta(S-S_0)$. We refine our theory by using the actual softness propagator in connecting the probability of rearranging, $P_R(S)$, with the overlap $q(S,t)$ (see Supplementary Information). For ease of calculation, we approximate $G(S,S_0,t)$ as a Gaussian distribution in $S$ and calculate its mean and variance as functions of $S_0$ and $t$ from simulated data. The resulting prediction for the overlap is shown in Fig.~3 (b). The agreement with the actual $q(t)$ is excellent, suggesting that an understanding of the time evolution of the softness field, or equivalently of the softness propagator, would suffice to understand the non-exponential decay of the overlap function.

Our results show that there is structure hidden in the disorder of glassy liquids. This structure can be quantified by softness, which controls glassy dynamics at temperatures below $T_0$. According to our analysis, simple Arrhenius relaxation for each softness, coupled with the time evolution of softness, leads to the observed slow, non-exponential relaxation dynamics of glassy liquids below $T_0$. Thus, our results suggest that the challenge of understanding glass transition dynamics can be reframed as the challenge of understanding the evolution of softness. 

SSS and EDC contributed equally to this work. We thank David Chandler, Kranthi Mandadapu, Daniel Sussman, Michele Parrinello, James Sethna, and David Wales for helpful discussions. This work was supported by the U.S. Department of Energy, Office of Basic Energy Sciences, Division of Materials Sciences and Engineering under Award DE-FG02-05ER46199 (S.S.S. and A.J.L.), and Harvard IACS Student Scholarship (E.D.C.). This work was partially supported by a grant from the Simons Foundation (\#305547 to A.J.L.).

\section{Methods}
\subsection{System information.}
We study a 10,000-particle Kob-Andersen model, a  80:20 binary LJ mixture~\cite{kob94} with parameters: $\sigma_{AA}=1.0$, $\sigma_{AB} = 0.8$, $\sigma_{BB} = 0.88$, $\epsilon_{AA} = 1.0$, $\epsilon_{AB}=1.5$, $\epsilon_{BB}=0.5$. Time is measured in units of $\tau = \sqrt{\epsilon_{AA}/\sigma_{AA}^2}$ and the Boltzmann constant is $k_B=1$. We cut off the LJ potential at $2.5\sigma_{AA}$ and smooth the potential so that force varies continuously. This mixture has been characterized extensively. In particular, we compare our predictions to  the measurements of the onset temperature in Keys \textit{et al.}~\cite{keys11}. Simulations were done using LAMMPS~\cite{plimpton95} in an NVT ensemble with a Nos\'e-Hoover thermostat and a timestep of $0.0025\tau$. We output states every $\tau$ and quench them to their nearest inherent structure using a combination of conjugate gradient and FIRE algorithms. Throughout this study we use inherent structure positions. However, qualitatively similar results can be obtained using time averaged positions. We study this system over the temperatures and number densities listed in Table \ref{tab:temperatures}.
\begin{table}
\centering
\caption{Number densities and temperatures studied.  Each column contains the temperatures studied for a given number density $\rho$.}
\label{tab:temperatures}
\medskip
\begin{tabular}{ccccc}
\hline
\hspace{0.5pc}$\rho$\hspace{0.5pc} & 1.15 & 1.20 & 1.25 & 1.30\\
\hline
T & 0.37 & 0.47 & 0.58 & 0.70\\
 & 0.42 & 0.51 & 0.61 & 0.75\\
 & 0.45 & 0.53 & 0.69 & 0.84\\
 & 0.52 & 0.56 & 0.76 & 0.92\\
 &  & 0.58 & 0.97 & 1.12\\
 &  & 0.70 & 0.97 & \\
\hline
\end{tabular}
\end{table}

\subsection{Identifying rearrangements.}
We adapt a method first proposed by Candelier \textit{et al.}\cite{candelier10,smessaert13}. A timescale $t_R=10\tau$ is chosen to be commensurate with the amount of time the system takes to complete a rearrangement. Then two time intervals are defined as $A=[t-t_R/2,t]$ and $B=[t,t+t_R/2]$. An indicator function can then be written as, 
\begin{equation}
p_{\text{hop}}(t) = \sqrt{\langle(\vec r_i - \langle \vec r_i\rangle_B)^2\rangle_A\langle(\vec r_i - \langle \vec r_i\rangle_A)^2\rangle_B}
\end{equation}
where $\langle\rangle_A$ and $\langle\rangle_B$ are averages over the intervals $A$ and $B$ respectively. $p_{\text{hop}}$ is large when the mean position of a particle changes appreciably. Otherwise, it is similar in magnitude to the variance in particle positions due to noise from the inherent structure calculation.

To find rearrangements we restrict our attention to events in which $p_{\text{hop}}$ exceeds a threshold of $0.05$, that is large compared to the scale of fluctuations in particle positions but small compared to the typical value of $p_{\text{hop}}$ during a rearrangement.
As discussed in the supplementary material, we define rearrangements to be those events with $p_{\text{hop}}^* > p_c= 0.2$. Changing this cutoff affects the results only quantitatively and manifests itself primarily as a shift in the energy scale, $\Delta E$ that is approximately logarithmic in the cutoff. This agrees with the observations of Keys \textit{et al.}~\cite{keys11} who saw a similar logarithmic shift in the energy scale governing rearrangements with the size of the rearrangements. 

Note that rearrangements defined using $p_{\text{hop}}$ result in particle displacements that follow a distribution that depends on the cutoff $p_c$ used. This $p_c$ dependence needs to be addressed when comparing the probability of rearrangement to the overlap function and its derivative, which are defined in terms of a length scale $a$. To do this, we multiply $P_R$ by a temperature-independent constant $c_a$, namely the fraction of rearrangements that displace particles by more than $a$.

\subsection{Computing softness.}
We have made two improvements that greatly increased the prediction accuracy for rearrangements compared to Ref.~\cite{cubukschoenholz15}. First, we identified rearrangements more carefully, as detailed above.  Second, we defined our training sets more carefully.  Each training set contains 6000 particles that rearrange in the next time step, each labeled with $r_i=1$, as well as 6000 particles that have not rearranged for a time $\tau_\alpha$ before the structure was calculated, each labeled with $r_i=0$.  These particles were chosen randomly from the set of all particles satisfying these conditions from MD simulations at a low temperature. Then, a training set of N particles can be written as $\left\{\left(\bm  F_1,r_1\right),...,\left(\bm F_N,r_N\right)\right\}$, where $\bm F_i$ = $\left\{F^1_i,...,F^M_i\right\}$  are the the $M$ structure functions that describe the local neighborhood of particle $i$~\cite{cubukschoenholz15}. We then use an SVM to find the hyperplane $\bm w\cdot \bm F - b = 0$ that separates the points with $r_i=1$ from those with $r_i=0$. 
This hyperplane is used on the rest of the data to reach the results reported.       

The SVM is trained, that is, the hyperplane is constructed, on the binary variable $r$ using 
the LIBSVM package~\cite{libsvm}. It is not possible to find a hyperplane that perfectly separates the two different classes. 
We use a penalty parameter $C$ and find the optimal hyperplane equation by minimizing
\begin{equation}
 \frac{1}{2} \bm w^T \cdot \bm w + C \sum_{i=1}^N \xi_i ,
\end{equation}
with the constraint
 $y_i\cdot \left(\bm w^T \cdot \bm F_i + b \right) \ge 1 - \xi_i$ and $\xi_i \ge 0$.
The $C$ parameter was chosen through cross-validation~\cite{cubukschoenholz15}. The hyperplane obtained from this training can be used to classify a new particle neighborhood, $\bm F_n$, as soft or hard. $\bm F_n$ is soft if $\bm w\cdot \bm F_n - b > 0$, and hard otherwise. The continuous variable softness is defined by  $S_n = \bm w\cdot \bm F_n - b$. Training a neural network to classify soft and hard particles, and using the output from the hidden layer of the neural network as softness, yields similar results. Here we use only the SVM approach. 



\bibliography{bibliography}

\begin{thebibliography}{24}%
\makeatletter
\providecommand \@ifxundefined [1]{%
 \@ifx{#1\undefined}
}%
\providecommand \@ifnum [1]{%
 \ifnum #1\expandafter \@firstoftwo
 \else \expandafter \@secondoftwo
 \fi
}%
\providecommand \@ifx [1]{%
 \ifx #1\expandafter \@firstoftwo
 \else \expandafter \@secondoftwo
 \fi
}%
\providecommand \natexlab [1]{#1}%
\providecommand \enquote  [1]{``#1''}%
\providecommand \bibnamefont  [1]{#1}%
\providecommand \bibfnamefont [1]{#1}%
\providecommand \citenamefont [1]{#1}%
\providecommand \href@noop [0]{\@secondoftwo}%
\providecommand \href [0]{\begingroup \@sanitize@url \@href}%
\providecommand \@href[1]{\@@startlink{#1}\@@href}%
\providecommand \@@href[1]{\endgroup#1\@@endlink}%
\providecommand \@sanitize@url [0]{\catcode `\\12\catcode `\$12\catcode
  `\&12\catcode `\#12\catcode `\^12\catcode `\_12\catcode `\%12\relax}%
\providecommand \@@startlink[1]{}%
\providecommand \@@endlink[0]{}%
\providecommand \url  [0]{\begingroup\@sanitize@url \@url }%
\providecommand \@url [1]{\endgroup\@href {#1}{\urlprefix }}%
\providecommand \urlprefix  [0]{URL }%
\providecommand \Eprint [0]{\href }%
\providecommand \doibase [0]{http://dx.doi.org/}%
\providecommand \selectlanguage [0]{\@gobble}%
\providecommand \bibinfo  [0]{\@secondoftwo}%
\providecommand \bibfield  [0]{\@secondoftwo}%
\providecommand \translation [1]{[#1]}%
\providecommand \BibitemOpen [0]{}%
\providecommand \bibitemStop [0]{}%
\providecommand \bibitemNoStop [0]{.\EOS\space}%
\providecommand \EOS [0]{\spacefactor3000\relax}%
\providecommand \BibitemShut  [1]{\csname bibitem#1\endcsname}%
\let\auto@bib@innerbib\@empty
\bibitem [{\citenamefont {Angell}(1995)}]{angell95}%
  \BibitemOpen
  \bibfield  {author} {\bibinfo {author} {\bibfnamefont {C.~A.}\ \bibnamefont
  {Angell}},\ }\href@noop {} {\bibfield  {journal} {\bibinfo  {journal}
  {Science}\ }\textbf {\bibinfo {volume} {267}},\ \bibinfo {pages} {1924}
  (\bibinfo {year} {1995})}\BibitemShut {NoStop}%
\bibitem [{\citenamefont {Sastry}\ \emph {et~al.}(1998)\citenamefont {Sastry},
  \citenamefont {Debenedetti},\ and\ \citenamefont {Stillinger}}]{sastry98}%
  \BibitemOpen
  \bibfield  {author} {\bibinfo {author} {\bibfnamefont {S.}~\bibnamefont
  {Sastry}}, \bibinfo {author} {\bibfnamefont {P.~G.}\ \bibnamefont
  {Debenedetti}}, \ and\ \bibinfo {author} {\bibfnamefont {F.~H.}\ \bibnamefont
  {Stillinger}},\ }\href@noop {} {\bibfield  {journal} {\bibinfo  {journal}
  {Nature}\ }\textbf {\bibinfo {volume} {393}},\ \bibinfo {pages} {554}
  (\bibinfo {year} {1998})}\BibitemShut {NoStop}%
\bibitem [{\citenamefont {Debenedetti}\ and\ \citenamefont
  {Stillinger}(2001)}]{debenedetti01}%
  \BibitemOpen
  \bibfield  {author} {\bibinfo {author} {\bibfnamefont {P.~G.}\ \bibnamefont
  {Debenedetti}}\ and\ \bibinfo {author} {\bibfnamefont {F.~H.}\ \bibnamefont
  {Stillinger}},\ }\href@noop {} {\bibfield  {journal} {\bibinfo  {journal}
  {Nature}\ }\textbf {\bibinfo {volume} {410}},\ \bibinfo {pages} {259}
  (\bibinfo {year} {2001})}\BibitemShut {NoStop}%
\bibitem [{\citenamefont {Parisi}\ and\ \citenamefont
  {Zamponi}(2010)}]{parisi10}%
  \BibitemOpen
  \bibfield  {author} {\bibinfo {author} {\bibfnamefont {G.}~\bibnamefont
  {Parisi}}\ and\ \bibinfo {author} {\bibfnamefont {F.}~\bibnamefont
  {Zamponi}},\ }\href@noop {} {\bibfield  {journal} {\bibinfo  {journal}
  {Reviews of Modern Physics}\ }\textbf {\bibinfo {volume} {82}},\ \bibinfo
  {pages} {789} (\bibinfo {year} {2010})}\BibitemShut {NoStop}%
\bibitem [{\citenamefont {Berthier}\ \emph {et~al.}(2011)\citenamefont
  {Berthier}, \citenamefont {Biroli}, \citenamefont {Bouchaud}, \citenamefont
  {Cipelletti},\ and\ \citenamefont {van Saarloos}}]{berthier11}%
  \BibitemOpen
  \bibfield  {author} {\bibinfo {author} {\bibfnamefont {L.}~\bibnamefont
  {Berthier}}, \bibinfo {author} {\bibfnamefont {G.}~\bibnamefont {Biroli}},
  \bibinfo {author} {\bibfnamefont {J.-P.}\ \bibnamefont {Bouchaud}}, \bibinfo
  {author} {\bibfnamefont {L.}~\bibnamefont {Cipelletti}}, \ and\ \bibinfo
  {author} {\bibfnamefont {W.}~\bibnamefont {van Saarloos}},\ }\href@noop {}
  {\emph {\bibinfo {title} {Dynamical Heterogeneities in Glasses, Colloids, and
  Granular Media}}}\ (\bibinfo  {publisher} {Oxford Scholarships Online},\
  \bibinfo {year} {2011})\BibitemShut {NoStop}%
\bibitem [{\citenamefont {Charbonneau}\ \emph {et~al.}(2014)\citenamefont
  {Charbonneau}, \citenamefont {Kurchan}, \citenamefont {Parisi}, \citenamefont
  {Urbani},\ and\ \citenamefont {Zamponi}}]{charbonneau14}%
  \BibitemOpen
  \bibfield  {author} {\bibinfo {author} {\bibfnamefont {P.}~\bibnamefont
  {Charbonneau}}, \bibinfo {author} {\bibfnamefont {J.}~\bibnamefont
  {Kurchan}}, \bibinfo {author} {\bibfnamefont {G.}~\bibnamefont {Parisi}},
  \bibinfo {author} {\bibfnamefont {P.}~\bibnamefont {Urbani}}, \ and\ \bibinfo
  {author} {\bibfnamefont {F.}~\bibnamefont {Zamponi}},\ }\href@noop {}
  {\bibfield  {journal} {\bibinfo  {journal} {J. Stat. Mech.}\ ,\ \bibinfo
  {pages} {10009}} (\bibinfo {year} {2014})}\BibitemShut {NoStop}%
\bibitem [{\citenamefont {Gilman}(1975)}]{gilman75}%
  \BibitemOpen
  \bibfield  {author} {\bibinfo {author} {\bibfnamefont {J.}~\bibnamefont
  {Gilman}},\ }\href@noop {} {\bibfield  {journal} {\bibinfo  {journal} {Phys.
  Today}\ }\textbf {\bibinfo {volume} {28}},\ \bibinfo {pages} {46} (\bibinfo
  {year} {1975})}\BibitemShut {NoStop}%
\bibitem [{\citenamefont {Berthier}\ and\ \citenamefont
  {Jack}(2007)}]{berthier07}%
  \BibitemOpen
  \bibfield  {author} {\bibinfo {author} {\bibfnamefont {L.}~\bibnamefont
  {Berthier}}\ and\ \bibinfo {author} {\bibfnamefont {R.~L.}\ \bibnamefont
  {Jack}},\ }\href@noop {} {\bibfield  {journal} {\bibinfo  {journal} {Phys.
  Rev. E}\ }\textbf {\bibinfo {volume} {76}},\ \bibinfo {pages} {041509}
  (\bibinfo {year} {2007})}\BibitemShut {NoStop}%
\bibitem [{\citenamefont {Royall}\ \emph {et~al.}(2008)\citenamefont {Royall},
  \citenamefont {Williams}, \citenamefont {Ohtsuka},\ and\ \citenamefont
  {Tanaka}}]{royall08}%
  \BibitemOpen
  \bibfield  {author} {\bibinfo {author} {\bibfnamefont {C.~P.}\ \bibnamefont
  {Royall}}, \bibinfo {author} {\bibfnamefont {S.~R.}\ \bibnamefont
  {Williams}}, \bibinfo {author} {\bibfnamefont {T.}~\bibnamefont {Ohtsuka}}, \
  and\ \bibinfo {author} {\bibfnamefont {H.}~\bibnamefont {Tanaka}},\
  }\href@noop {} {\bibfield  {journal} {\bibinfo  {journal} {Nature materials}\
  }\textbf {\bibinfo {volume} {7}},\ \bibinfo {pages} {556} (\bibinfo {year}
  {2008})}\BibitemShut {NoStop}%
\bibitem [{\citenamefont {Manning}\ and\ \citenamefont
  {Liu}(2011)}]{manning11}%
  \BibitemOpen
  \bibfield  {author} {\bibinfo {author} {\bibfnamefont {M.~L.}\ \bibnamefont
  {Manning}}\ and\ \bibinfo {author} {\bibfnamefont {A.~J.}\ \bibnamefont
  {Liu}},\ }\href@noop {} {\bibfield  {journal} {\bibinfo  {journal} {Phys.
  Rev. Lett.}\ }\textbf {\bibinfo {volume} {107}},\ \bibinfo {pages} {108302}
  (\bibinfo {year} {2011})}\BibitemShut {NoStop}%
\bibitem [{\citenamefont {Jack}\ \emph {et~al.}(2014)\citenamefont {Jack},
  \citenamefont {Dunleavy},\ and\ \citenamefont {Royall}}]{jack14}%
  \BibitemOpen
  \bibfield  {author} {\bibinfo {author} {\bibfnamefont {R.~L.}\ \bibnamefont
  {Jack}}, \bibinfo {author} {\bibfnamefont {A.~J.}\ \bibnamefont {Dunleavy}},
  \ and\ \bibinfo {author} {\bibfnamefont {C.~P.}\ \bibnamefont {Royall}},\
  }\href@noop {} {\bibfield  {journal} {\bibinfo  {journal} {Phys. Rev. Lett.}\
  }\textbf {\bibinfo {volume} {113}},\ \bibinfo {pages} {095703} (\bibinfo
  {year} {2014})}\BibitemShut {NoStop}%
\bibitem [{\citenamefont {Cubuk}\ \emph {et~al.}(2015)\citenamefont {Cubuk},
  \citenamefont {Schoenholz}, \citenamefont {Rieser}, \citenamefont {Malone},
  \citenamefont {Rottler}, \citenamefont {Durian}, \citenamefont {Kaxiras},\
  and\ \citenamefont {Liu}}]{cubukschoenholz15}%
  \BibitemOpen
  \bibfield  {author} {\bibinfo {author} {\bibfnamefont {E.~D.}\ \bibnamefont
  {Cubuk}}, \bibinfo {author} {\bibfnamefont {S.~S.}\ \bibnamefont
  {Schoenholz}}, \bibinfo {author} {\bibfnamefont {J.~M.}\ \bibnamefont
  {Rieser}}, \bibinfo {author} {\bibfnamefont {B.~D.}\ \bibnamefont {Malone}},
  \bibinfo {author} {\bibfnamefont {J.}~\bibnamefont {Rottler}}, \bibinfo
  {author} {\bibfnamefont {D.~J.}\ \bibnamefont {Durian}}, \bibinfo {author}
  {\bibfnamefont {E.}~\bibnamefont {Kaxiras}}, \ and\ \bibinfo {author}
  {\bibfnamefont {A.~J.}\ \bibnamefont {Liu}},\ }\href@noop {} {\bibfield
  {journal} {\bibinfo  {journal} {Phys. Rev. Lett.}\ }\textbf {\bibinfo
  {volume} {114}},\ \bibinfo {pages} {108001} (\bibinfo {year}
  {2015})}\BibitemShut {NoStop}%
\bibitem [{\citenamefont {Behler}\ and\ \citenamefont
  {Parrinello}(2007)}]{behler07}%
  \BibitemOpen
  \bibfield  {author} {\bibinfo {author} {\bibfnamefont {J.}~\bibnamefont
  {Behler}}\ and\ \bibinfo {author} {\bibfnamefont {M.}~\bibnamefont
  {Parrinello}},\ }\href@noop {} {\bibfield  {journal} {\bibinfo  {journal}
  {Phys. Rev. Lett.}\ }\textbf {\bibinfo {volume} {98}},\ \bibinfo {pages}
  {146401} (\bibinfo {year} {2007})}\BibitemShut {NoStop}%
\bibitem [{\citenamefont {Chang}\ and\ \citenamefont {Lin}(2011)}]{libsvm}%
  \BibitemOpen
  \bibfield  {author} {\bibinfo {author} {\bibfnamefont {C.-C.}\ \bibnamefont
  {Chang}}\ and\ \bibinfo {author} {\bibfnamefont {C.-J.}\ \bibnamefont
  {Lin}},\ }\href@noop {} {\bibfield  {journal} {\bibinfo  {journal} {ACM
  Transactions on Intelligent Systems and Technology}\ }\textbf {\bibinfo
  {volume} {2}},\ \bibinfo {pages} {27:1} (\bibinfo {year} {2011})}\BibitemShut
  {NoStop}%
\bibitem [{\citenamefont {Cortes}\ and\ \citenamefont {Vapnik}(1995)}]{SVM}%
  \BibitemOpen
  \bibfield  {author} {\bibinfo {author} {\bibfnamefont {C.}~\bibnamefont
  {Cortes}}\ and\ \bibinfo {author} {\bibfnamefont {V.}~\bibnamefont
  {Vapnik}},\ }\href@noop {} {\bibfield  {journal} {\bibinfo  {journal} {Mach.
  Learn.}\ }\textbf {\bibinfo {volume} {20}},\ \bibinfo {pages} {273} (\bibinfo
  {year} {1995})}\BibitemShut {NoStop}%
\bibitem [{\citenamefont {Kob}\ and\ \citenamefont {Andersen}(1994)}]{kob94}%
  \BibitemOpen
  \bibfield  {author} {\bibinfo {author} {\bibfnamefont {W.}~\bibnamefont
  {Kob}}\ and\ \bibinfo {author} {\bibfnamefont {H.~C.}\ \bibnamefont
  {Andersen}},\ }\href@noop {} {\bibfield  {journal} {\bibinfo  {journal}
  {Phys. Rev. Lett.}\ }\textbf {\bibinfo {volume} {73}},\ \bibinfo {pages}
  {1376} (\bibinfo {year} {1994})}\BibitemShut {NoStop}%
\bibitem [{\citenamefont {Berthier}\ and\ \citenamefont
  {Biroli}(2011)}]{berthier11_2}%
  \BibitemOpen
  \bibfield  {author} {\bibinfo {author} {\bibfnamefont {L.}~\bibnamefont
  {Berthier}}\ and\ \bibinfo {author} {\bibfnamefont {G.}~\bibnamefont
  {Biroli}},\ }\href@noop {} {\bibfield  {journal} {\bibinfo  {journal}
  {Reviews of Modern Physics}\ }\textbf {\bibinfo {volume} {83}},\ \bibinfo
  {pages} {587} (\bibinfo {year} {2011})}\BibitemShut {NoStop}%
\bibitem [{\citenamefont {Percus}\ and\ \citenamefont
  {Yevick}(1958)}]{percus58}%
  \BibitemOpen
  \bibfield  {author} {\bibinfo {author} {\bibfnamefont {J.~K.}\ \bibnamefont
  {Percus}}\ and\ \bibinfo {author} {\bibfnamefont {G.~J.}\ \bibnamefont
  {Yevick}},\ }\href@noop {} {\bibfield  {journal} {\bibinfo  {journal} {Phys.
  Rev.}\ }\textbf {\bibinfo {volume} {110}},\ \bibinfo {pages} {1} (\bibinfo
  {year} {1958})}\BibitemShut {NoStop}%
\bibitem [{\citenamefont {Smessaert}\ and\ \citenamefont
  {Rottler}(2014)}]{smessaert14}%
  \BibitemOpen
  \bibfield  {author} {\bibinfo {author} {\bibfnamefont {A.}~\bibnamefont
  {Smessaert}}\ and\ \bibinfo {author} {\bibfnamefont {J.}~\bibnamefont
  {Rottler}},\ }\href {\doibase 10.1039/C4SM01438C} {\bibfield  {journal}
  {\bibinfo  {journal} {Soft Matter}\ }\textbf {\bibinfo {volume} {10}},\
  \bibinfo {pages} {8533} (\bibinfo {year} {2014})}\BibitemShut {NoStop}%
\bibitem [{\citenamefont {Keys}\ \emph {et~al.}(2011)\citenamefont {Keys},
  \citenamefont {Hedges}, \citenamefont {Garrahan}, \citenamefont {Glotzer},\
  and\ \citenamefont {Chandler}}]{keys11}%
  \BibitemOpen
  \bibfield  {author} {\bibinfo {author} {\bibfnamefont {A.~S.}\ \bibnamefont
  {Keys}}, \bibinfo {author} {\bibfnamefont {L.~O.}\ \bibnamefont {Hedges}},
  \bibinfo {author} {\bibfnamefont {J.~P.}\ \bibnamefont {Garrahan}}, \bibinfo
  {author} {\bibfnamefont {S.~C.}\ \bibnamefont {Glotzer}}, \ and\ \bibinfo
  {author} {\bibfnamefont {D.}~\bibnamefont {Chandler}},\ }\href@noop {}
  {\bibfield  {journal} {\bibinfo  {journal} {Phys. Rev. X.}\ }\textbf
  {\bibinfo {volume} {1}},\ \bibinfo {pages} {021013} (\bibinfo {year}
  {2011})}\BibitemShut {NoStop}%
\bibitem [{\citenamefont {Keys}\ \emph {et~al.}(2007)\citenamefont {Keys},
  \citenamefont {Abate}, \citenamefont {Glotzer},\ and\ \citenamefont
  {Durian}}]{keys07}%
  \BibitemOpen
  \bibfield  {author} {\bibinfo {author} {\bibfnamefont {A.~S.}\ \bibnamefont
  {Keys}}, \bibinfo {author} {\bibfnamefont {A.~R.}\ \bibnamefont {Abate}},
  \bibinfo {author} {\bibfnamefont {S.~C.}\ \bibnamefont {Glotzer}}, \ and\
  \bibinfo {author} {\bibfnamefont {D.~J.}\ \bibnamefont {Durian}},\
  }\href@noop {} {\bibfield  {journal} {\bibinfo  {journal} {Nat. Phys.}\
  }\textbf {\bibinfo {volume} {3}},\ \bibinfo {pages} {260} (\bibinfo {year}
  {2007})}\BibitemShut {NoStop}%
\bibitem [{\citenamefont {Plimpton}(1995)}]{plimpton95}%
  \BibitemOpen
  \bibfield  {author} {\bibinfo {author} {\bibfnamefont {S.}~\bibnamefont
  {Plimpton}},\ }\href@noop {} {\bibfield  {journal} {\bibinfo  {journal} {J.
  Comp. Phys.}\ }\textbf {\bibinfo {volume} {117}},\ \bibinfo {pages} {1}
  (\bibinfo {year} {1995})}\BibitemShut {NoStop}%
\bibitem [{\citenamefont {Candelier}\ \emph {et~al.}(2010)\citenamefont
  {Candelier}, \citenamefont {Widmer-Cooper}, \citenamefont {Kummerfeld},
  \citenamefont {Dauchot}, \citenamefont {Biroli}, \citenamefont {Harrowell},\
  and\ \citenamefont {Reichman}}]{candelier10}%
  \BibitemOpen
  \bibfield  {author} {\bibinfo {author} {\bibfnamefont {R.}~\bibnamefont
  {Candelier}}, \bibinfo {author} {\bibfnamefont {A.}~\bibnamefont
  {Widmer-Cooper}}, \bibinfo {author} {\bibfnamefont {J.~K.}\ \bibnamefont
  {Kummerfeld}}, \bibinfo {author} {\bibfnamefont {O.}~\bibnamefont {Dauchot}},
  \bibinfo {author} {\bibfnamefont {G.}~\bibnamefont {Biroli}}, \bibinfo
  {author} {\bibfnamefont {P.}~\bibnamefont {Harrowell}}, \ and\ \bibinfo
  {author} {\bibfnamefont {D.~R.}\ \bibnamefont {Reichman}},\ }\href@noop {}
  {\bibfield  {journal} {\bibinfo  {journal} {Phys. Rev. Lett.}\ }\textbf
  {\bibinfo {volume} {105}},\ \bibinfo {pages} {135702} (\bibinfo {year}
  {2010})}\BibitemShut {NoStop}%
\bibitem [{\citenamefont {Smessaert}\ and\ \citenamefont
  {Rottler}(2013)}]{smessaert13}%
  \BibitemOpen
  \bibfield  {author} {\bibinfo {author} {\bibfnamefont {A.}~\bibnamefont
  {Smessaert}}\ and\ \bibinfo {author} {\bibfnamefont {J.}~\bibnamefont
  {Rottler}},\ }\href@noop {} {\bibfield  {journal} {\bibinfo  {journal}
  {Physical Review E}\ }\textbf {\bibinfo {volume} {88}},\ \bibinfo {pages}
  {022314} (\bibinfo {year} {2013})}\BibitemShut {NoStop}%
\end{thebibliography}%



\end{document}